\documentstyle[11pt,epsf]{article}
\begin{document}
\thispagestyle{empty}
\rightline{UOSTP-00-102}
\rightline{{\tt hep-th/0005253}}

\

\def\tr{{\rm tr}\,} \newcommand{\beq}{\begin{equation}}
\newcommand{\eeq}{\end{equation}} \newcommand{\beqn}{\begin{eqnarray}}
\newcommand{\eeqn}{\end{eqnarray}} \newcommand{\bde}{{\bf e}}
\newcommand{\balpha}{{\mbox{\boldmath $\alpha$}}}
\newcommand{\bsalpha}{{\mbox{\boldmath $\scriptstyle\alpha$}}}
\newcommand{\betabf}{{\mbox{\boldmath $\beta$}}}
\newcommand{\bgamma}{{\mbox{\boldmath $\gamma$}}}
\newcommand{\bbeta}{{\mbox{\boldmath $\scriptstyle\beta$}}}
\newcommand{\lambdabf}{{\mbox{\boldmath $\lambda$}}}
\newcommand{\bphi}{{\mbox{\boldmath $\phi$}}}
\newcommand{\bslambda}{{\mbox{\boldmath $\scriptstyle\lambda$}}}
\newcommand{\ggg}{{\boldmath \gamma}} \newcommand{\ddd}{{\boldmath
\delta}} \newcommand{\mmm}{{\boldmath \mu}}
\newcommand{\nnn}{{\boldmath \nu}}
\newcommand{\diag}{{\rm diag}}
\newcommand{\bra}[1]{\langle {#1}|}
\newcommand{\ket}[1]{|{#1}\rangle}
\newcommand{\sn}{{\rm sn}}
\newcommand{\cn}{{\rm cn}}
\newcommand{\dn}{{\rm dn}}
\newcommand{\tA}{{\tilde{A}}}
\newcommand{\tphi}{{\tilde\phi}}
\newcommand{\bpartial}{{\bar\partial}}
\newcommand{\br}{{{\bf r}}}
\newcommand{\bx}{{{\bf x}}}
\newcommand{\bk}{{{\bf k}}}
\newcommand{\bq}{{{\bf q}}}
\newcommand{\bQ}{{{\bf Q}}}
\newcommand{\bp}{{{\bf p}}}
\newcommand{\bP}{{{\bf P}}}
\newcommand{\thet}{{{\theta}}}
\renewcommand{\thefootnote}{\fnsymbol{footnote}}
\

\vskip 0cm
\centerline{\Large \bf 
Exact Wavefunctions in
a Noncommutative Field Theory\footnote{This work 
is supported
in part by KOSEF 1998 Interdisciplinary Research Grant
98-07-02-07-01-5 and by BK21 Project of
Ministry of Education.
}}

\vskip .2cm

\vskip 1.2cm
\centerline{ Dongsu Bak,${}^a\!\!$
\footnote{Electronic Mail: dsbak@mach.uos.ac.kr}
 Sung Ku Kim,${}^b$
Kwang-Sup Soh${}^c$ and Jae Hyung Yee${}^d$
}
\vskip 7mm
\centerline{Physics Department, 
University of Seoul, Seoul 130-743 Korea${}^a$}
\vskip0.3cm
\centerline{Physics Department, Ewha Women's University,
Seoul 120-750 Korea${}^b$}
\vskip0.3cm
\centerline{Physics Department, Seoul National University,
Seoul 151-742 Korea${}^c$}
\vskip0.3cm
\centerline{Institute of Physics and Applied Physics, 
Yonsei University,
Seoul 120-749 Korea${}^d$}
\vskip0.4cm
\vskip 3mm

\vskip 1.2cm
\begin{quote}
{
We consider the nonrelativistic field theory with a 
quartic
interaction 
on a noncommutative 
plane. We compute the $2\!\rightarrow\!2$ scattering amplitude 
within perturbative analysis to all orders
and identify the 
beta function and the running of the coupling constant. 
Since the theory admits an equivalent description via
the N particle Schr\"odinger equation, we regain 
the scattering amplitude by finding an 
exact scattering wavefunction of the two body equation. 
The wave function for the bound state is also 
identified.  These wave functions 
 unusually have  two center positions in the relative 
coordinates. 
The separation of the centers is in the transverse direction
of the total momentum and grows linearly with 
the noncommutativity 
scale and the total momentum, exhibiting the stringy nature of
the noncommutative field theory.
}
\end{quote}


\newpage

As shown recently, 
quantum field theories in noncommutative spacetime
naturally arise as a decoupling limit of the 
worldvolume dynamics of D-branes in a constant NS-NS 
two form background\cite{seiberg}. 
The 
dynamical effects of the 
noncommutative geometries mainly 
in the classical context have been 
investigated\cite{seiberg,connes,douglas,itzhaki,nekrasov,maldacena}.
The quantum aspects of the noncommutative field theories
are also  pursued via perturbative analysis over diverse 
models\cite{minwalla,filk,gomis,matusis}, but the understanding 
is still partial.
 
 In this note, we shall consider
the nonrelativistic system with a contact 
interaction\cite{jackiw,bergman} 
on a noncommutative plane. Performing the perturbative 
analysis, we obtain the $2\!\rightarrow\! 2$ scattering 
amplitude exactly. We prove that the theory is 
renormalizable in the two particle sector 
and identify the running of the coupling constant
to all orders. The theory admits an equivalent description via
the N-particle
Schr\"odinger equation. We find the scattering and the bound state
wavefunctions exactly and show 
that the bound state energy and the scattering amplitude  
agree
with those from the perturbation theory. It is observed that 
the exact two particle wavefunctions 
have two centers, whose separation is
in the transverse direction of the total 
momentum. The separation grows linearly as the noncommutative scale 
and the magnitude of the total momentum, 
exhibiting the stringy nature of
the noncommutative field theory\cite{susskind}.

We shall consider a nonrelativistic  
scalar field theory on a noncommutative plane described by the 
Lagrangian,
\begin{equation}
L=\int d^2x\left( i\,\psi^\dagger \partial_t \psi +
{1\over 2} \psi^\dagger  \nabla^2
 \psi 
 -{v\over 4} \psi^\dagger * \psi^\dagger * \psi * \psi
\right),
\label{lag}
\end{equation} 
where the $*$-product (Moyal product) is defined by
\begin{equation}
a(\bx)* b(\bx)\equiv \Bigl(e^{{i\over 2}\theta^{ij}\partial_i 
\partial'_j} a(\bx) b(\bx')\Bigr){\Big\vert}_{\bx=\bx'}
\label{star}
\end{equation}
with $\theta^{ij}(\,=\thet \epsilon^{ij})$ being 
antisymmetric. 
Classically, the system in the ordinary spacetime 
possesses 
the scale invariance as well as the familiar Galileo 
symmetry. The scale invariance is broken quantum mechanically 
due to short distance singularities of the contact 
interaction, producing anomalies. In the noncommutative 
case, the full Galileo invariance is lost  but the rotational and 
translational
symmetries remain because the Moyal 
product is not covariant under only the boost operations. 
As a consequence, relative degrees of motion do not in general 
decouple  from total translational motion of the system. But 
the energy and momentum tensor is defined because the time and 
spatial translations are symmetries of the system.
The breaking of the scale invariance in the noncommutative case is 
two fold. It is from both the quantum effect and the 
explicit scale dependence of the Moyal product. The global $U(1)$ 
invariance under $\psi\rightarrow e^{i\alpha}\psi$  
 persists in the noncommutative case and the number
 operator $N=\int d^2 x\, \psi^\dagger\psi$ is still 
conserved.  

We shall quantize the system not by the path integral approach but 
by the canonical quantization methods imposing the canonical
commutation 
relation  
\begin{eqnarray}
[\psi(\bx),\psi^\dagger(\bx')]=\delta(\bx-\bx')\,.
\label{etcr}
\end{eqnarray}
The Hamiltonian is given by
\begin{eqnarray}
H=H_0 + V=\int d^2x\left(
-{1\over 2} \psi^\dagger  \nabla^2
 \psi 
 +{v\over 4} \psi^\dagger * \psi^\dagger * \psi * \psi
\right)
\label{hamiltonian}
\end{eqnarray} 
where $H_0$ and $V$ denote, respectively, the free Hamiltonian
 and the contact interaction term.
The field in the Schr\"odinger picture can be expanded 
in its Fourier components as
\begin{eqnarray}
\psi(\bx)=\int {d^2k\over (2\pi)^2} a(\bk) e^{i\bk\cdot\bx}\,,
\label{fourier}
\end{eqnarray} 
and we shall define a vacuum state $|0\rangle$ 
by $a(\bk)|0\rangle=0$.
Then the full two point Green function $\langle 0|T(\psi(\bx,t)
\psi^\dagger(\bx',t'))|0\rangle$ 
can be computed as follows.  
First notice 
that  $\psi(\bx,t)=e^{iHt}\psi(\bx)
e^{-iHt}$ and  $e^{-iHt}=e^{-iH_0 t} 
\, T(\exp{[-i\int_0^tdt' V_I (t')}])$, where 
interaction picture 
operators are defined by $O_I (t)=e^{iH_0t}O
e^{-iH_0t}$. 
Further using the facts  
$\langle 0| [\psi_I(\bx,t),V_I (t_1) \cdots V_I (t_n)]\, =0$
and $\langle 0| V_I (t)=0$,
one finds that 
$\langle 0|T(\psi(\bx,t)\psi^\dagger(\bx',t'))|0\rangle =
\langle 0|T(\psi_I(\bx,t)
\psi_I^\dagger(\bx',t'))|0\rangle$. Namely, 
the full two point Green function is the same as that of 
the free Schr\"odinger field.
The expression for the full two point Green function reads 
explicitly
\begin{eqnarray}
\langle 0|T(\psi(\bx,t)
\psi^\dagger(\bx',t'))|0\rangle 
= \int {d^2 k d\omega\over (2\pi)^3} 
{i\over \omega -{k^2\over 2}+i\epsilon} e^{-i\omega 
(t-t')+i\bk\cdot(\bx-\bx')}
\,.
\label{twopoint}
\end{eqnarray} 
 This implies 
that the full propagator is not corrected 
perturbatively, for example, by the tadpole Feynman diagram. 
In other words, 
the tadpole diagram is absent 
in the perturbative scheme defined 
below\footnote{The full Green function 
 here is hence different from that in 
Ref.~\cite{gomis} where the authors used the path integral 
quantization methods, in which the ordering prescribed in the
Hamiltonian (\ref{hamiltonian}) is not implemented.}.

{\large Perturbation Theory}
 
Our first goal is to calculate the $2\!\rightarrow\! 2$ scattering 
amplitude within perturbative analysis 
and to compare this with the result from two particle
Schr\"odinger equation. 
We shall adopt the perturbation theory defined by the canonical quantization
methods. 
The propagator of the bosonic field
is given by 
\begin{eqnarray}
D(k_0,\bk)=
{i\over k_0 -{k^2\over 2}+i\epsilon} \,.
\label{propagator}
\end{eqnarray} 
The denominator is linear in energy, resulting in propagation
that is only forward in time. For this reason, the number of 
nonvanishing Feynman diagrams is reduced a lot compared to
the relativistic version of the theory\cite{bergman}. 
The identification of the interaction vertex
\begin{eqnarray}
\Gamma_0\left(\bk_1\!+\!\bk_2,{\bk_1\!-\!\bk_2\over 2};
\bk_3\!+\!\bk_4,{\bk_3\!-\!\bk_4\over 2}\right)=-
iv_0\cos(k_1 \wedge k_2) 
\cos(k_3\wedge k_4)\,, 
\label{vertex}
\end{eqnarray} 
is straightforward where  $v_0$ denotes the bare 
coupling constant and $k\wedge k'
\equiv {\theta\over 2}\epsilon^{ij}k_i k'_j$.  
This also agrees with the expression in 
Ref.~\cite{gomis}. The propagator and vertex are presented 
diagrammatically in Fig.~1. 
\begin{figure}[tb]
\epsfxsize=2.7in
\hspace{.8in}\epsffile{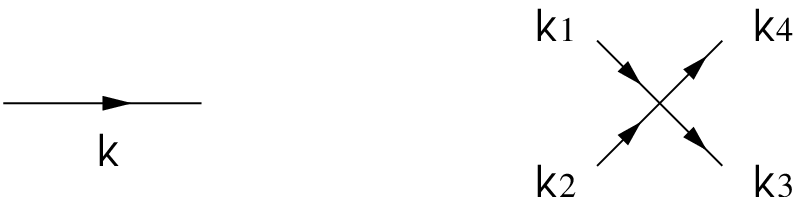}
\\
{\small Figure~1:~The Feynman diagrams for the 
 propagator and the vertex.}
\end{figure}
At this level, 
setting $\theta$ to zero,
one should recover the corresponding propagator and 
vertex of the theory in the ordinary plane.

For the scattering amplitude, we compute the 
on shell four point 
 function. The one loop bubble Feynman diagram is described in 
Fig.~2 and all the 1PI Feynman diagrams for the four point  function
are also depicted.
Apart from an energy and momentum conserving delta function, the one loop
bubble diagram is obtained evaluating the following expression,
\begin{eqnarray}
\Gamma_{\rm b}={1\over 2}\int d^2 q dq_0\,
\Gamma_0(\bP, \bk;\bP, \bq)\,\Gamma_0(\bP, \bq;\bP, \bk')
\,D\left({E_{\rm t}\over 2}\! +\!q_0, {{\bP\over 2}\!+\!\bq}\right) 
D\left({E_{\rm t}\over 2}\!-\!q_0, {\bP\over 2}\!-\!\bq\right)
\label{oneloop}
\end{eqnarray} 
where $E_{\rm t}$ denotes the total energy and 
the factor a half in front of the integral 
comes from the symmetry consideration. Once $q_0$
integration is performed, we are left with 
the UV divergent integral with respect to
$\bq$. This
will be regulated by introducing a large momentum cut-off 
$\Lambda$.
\begin{figure}[hb]
\epsfxsize=5in
\hspace{.8in}\epsffile{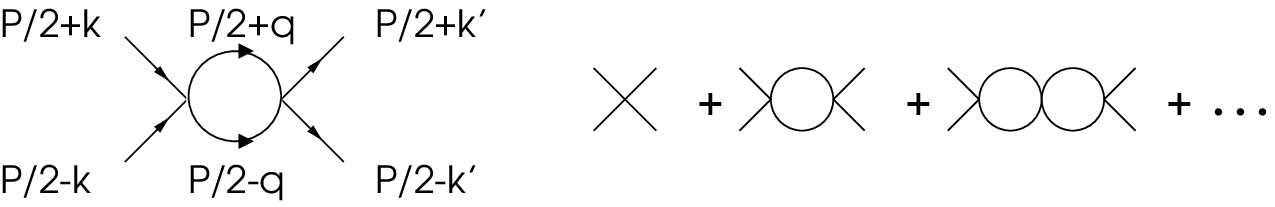}
\\
{\small Figure~2:~The bubble diagram and 
all the 1PI diagrams for the four point function.}
\end{figure} 
The contribution 
of the bubble diagram then reads\cite{gomis},
\begin{eqnarray}
\Gamma_{\rm b}&=&{i v_0^2\over 8\pi} \,
\cos(k\wedge P) 
\cos(k'\wedge P)
\int {d^2 q\over 2\pi}{1+ \cos(2 q\wedge P)\over
q^2-k^2-i\epsilon}\nonumber\\
&=& {i v_0^2\over 8\pi} \,
\cos(k\wedge P) 
\cos(k'\wedge P)
 \left[Z\left({\Lambda/k}\right)+ 
K_0 (-i\theta k P)\right]
\label{oneloopa} 
\end{eqnarray}
where $K_0(x)$ is the Bessel function of the imaginary argument
and $Z(x)\equiv  \ln{x}+{i\pi\over 2}$. If one set $\theta$ to zero
before sending the cut-off $\Lambda$ large, one would recover the 
one loop amplitude 
of the ordinary field theory, in which $Z+K_0$  above 
is replaced by $2 Z(\Lambda/k)$. 
Taking instead the $\theta\!\rightarrow\! 0$ limit 
of (\ref{oneloopa}), the amplitude of the ordinary 
field theory is not obtained, but the expression 
becomes singular. In other words, 
the $\theta\!\rightarrow\! 0$ limit does not commute with the 
$\Lambda\! \rightarrow\! \infty$ limit. 
Since it is fairly clear how to treat the case $\theta P=0$,
we shall below restrict our considerations to the case where
$\theta P$ is nonvanishing.
The double bubble,
the triple bubble diagrams and so on, can also be computed similarly.
Summing
these to all loops, one obtains the  full 
four point  
function as
\begin{eqnarray}
\Gamma ={-i \cos(k\wedge P) 
\cos(k'\wedge P)
 \over {1\over v_0}+{1\over 8\pi}
\left[ Z(\Lambda/k)+ 
K_0 (-i\theta k P)\right]}
\label{fullfour} 
\end{eqnarray}
A few comments are in order. For small  $x$,
$K_0 (-ix)$ can be expanded as $ K_0 (-ix)= 
Z(2/(\gamma x))+ 
O(x^2)$ with $\gamma$ being the Euler's constant.
Hence, one may see that the UV scale $\Lambda$ and the IR scale 
$\theta P$ combine in the four point function, 
replacing $Z(\Lambda/k)+ 
K_0 (-i\theta k P)$ by $2Z\left((2\Lambda)^{1\over 2}/
(\gamma\theta P k^2)^{1\over 2}\right)$ for small $\theta P$.
The renormalization  is achieved by redefining coupling 
constant $v_0$,
\begin{eqnarray}
{1\over v(\mu)}= {1\over v_0} + 
{1\over 8\pi}\ln\left({\Lambda\over \mu}\right)
\label{renormalization} 
\end{eqnarray}
where $\mu$ is an arbitrary renormalization 
scale\footnote{For the consistency of this relation, the bare coupling 
constant should be negative\cite{bergman}}.
Alternatively, the renormalization can be achieved by adding a
counterterm, $c_0\,V$, to the Lagrangian order by 
order. 
This proves that the theory is 
renormalizable in the two particle sector. 
The exact beta 
function controlling the running of the coupling constant
is found to be $\beta(v(\mu))\equiv \mu{\partial v(\mu)\over \partial 
\mu }={v^2(\mu)\over 8\pi}$, which is a half of the beta function
arising in the 
corresponding ordinary field theory. 
In terms of the the renormalized coupling constant, the 
$2\!\rightarrow\! 2$ scattering
amplitude is 
\begin{eqnarray}
A(\bk,\bk';\bP)=-{1\over 4\sqrt{\pi k}}\,\,{
\cos(k\wedge P) 
\cos(k'\wedge P)
\over {1\over v(\mu)}+{1\over 8\pi}
[ Z(\mu/k)+ 
K_0 (-i\theta k P)]}
\label{amplitude} 
\end{eqnarray}
where an appropriate kinematical factor is included. Later,
we shall compare this with the result from two body 
Schr\"odinger equation.  As said earlier, the amplitude 
depends upon
the total momentum $P$ signalling that the boost operation is no 
longer a symmetry of the system. The imaginary 
poles of the 
scattering amplitude indicate that there are bound states. 
The bound state 
energy $E_B=-\epsilon_B\, (\epsilon_B\geq 0)$ is 
obtained by solving
$1=-{v(\mu)\over 8\pi}
[ Z(-i\mu/\sqrt{\epsilon_B})+ K_0 (\theta \sqrt{\epsilon_B} P)]$ 
and there is a 
single bound state for given $v(\mu)$ since $Z\!+\!K_0$ 
is a monotonic
 function of $\epsilon_B$ covering the range 
$(-\infty,\infty)$.
There are also poles with real $k$ that correspond to 
resonant states. These states were absent in the case of the 
ordinary field theory.
 Further details will be investigated later in 
the quantum mechanical setting. 

{\large Two Particle Schr\"odinger Equation}

As mentioned before, the number operator  is conserved 
in the Schr\"odinger system. The eigenvalue of 
the number operators can be  shown to be a nonnegative integer 
that counts particle number.  One may then 
construct two particle Schr\"odinger equation by the following 
manner. First note that the operator Schr\"odinger 
equation is given by
\begin{eqnarray}
&&i{\partial\over \partial t}\psi(\br)=[\psi(\br), H]\nonumber\\
&&=-{1\over 2}\nabla^2 \psi(\br)+ {v\over 4}
\int d^2x\left(  \delta(\br-\bx)* \psi^\dagger(\bx)+ \psi^\dagger(\bx)* 
\delta(\br-\bx)\right) 
* \psi(\bx) * \psi(\bx)
\label{opeq}
\end{eqnarray} 
where the time argument of the Schr\"odinger field operator is 
suppressed for simplicity. The two particle wavefunction may be  
constructed by projecting a generic state $|\Phi\rangle$ to two 
particle sector, i.e. 
$\phi(\br,\br')\!=\!\langle 0|\psi(\br,t)\psi(\br',t)|\Phi\rangle$. 
Using the operator Schr\"odinger equation and 
evaluating 
$i\dot\phi(\br,\br')$,
one is led to the two particle Schr\"odinger equation,
\begin{eqnarray}
i\dot{\phi}(\br,\br')\!=\!-
{1\over 2}\!\left(\nabla^2\!+\!{\nabla'}^2\right)
\phi(\br,\br')\!+\! {v\over 4}
\int d^2x\left[  \delta(\br\!\!-\!\!\bx)\!*\!  
\delta(\br'\!\!-\!\!\bx)\!+\! 
\delta(\br'\!\!-\!\!\bx)\!*\! 
\delta(\br\!\!-\!\!\bx)\right]\!*\! 
\phi_*(\bx,\!\bx)
\label{twobody}
\end{eqnarray}  
where $\phi_*(\bx,\!\bx')=e^{{i\over 2}\theta^{ij}\partial_i
\partial'_j} \phi(\bx,\bx')$.
For the  momentum space representation, we  define 
$\phi(\bQ, \bq)=\int d^2 r d^2 r' 
e^{-{i\over 2}\bQ\cdot (\br+\br')-i\bq\cdot(\br-\br')}\phi(\br,\br')$.
The equation in the momentum space is then given by
\begin{eqnarray}
i\dot{\phi}(\bQ,\bq)\!=\left({1\over 4}{Q^2}+q^2\right)
\phi(\bQ,\bq)\!+\! 
{v_0 \cos(q\wedge Q)\over 8\pi^2}
\int d^2 q' \cos(q'\wedge Q)
\phi(\bQ,\bq')\,.
\label{twobodya}
\end{eqnarray}  
Setting 
$\phi(\bQ,\bq)= \delta(\bQ\!-\!\bP)\varphi(\bP,\bq) 
e^{-i({1\over 4}{P^2}+E_r)t}$,
the 
Schr\"odinger equation is reduced to
\begin{eqnarray}
(E_r-q^2)\varphi(\bP,\bq)=
{v_0 \cos(q\wedge P)\over 8\pi^2}
\int d^2 q' \cos(q'\wedge P)
\varphi(\bP,\bq')\,.
\label{twobodyb}
\end{eqnarray}  
For the bound state with $E_r\!=\! -\epsilon_B
$, 
the equation takes a form
\begin{eqnarray}
\varphi(\bP,\bq)=
{ \cos(q\wedge P)\over q^2+ \epsilon_B} C(\bP)\,.
\label{relative}
\end{eqnarray}  
with $C(\bP)=
-{v_0\over 8\pi^2}\int d^2 q' \cos(q'\wedge P)
\varphi(\bP,\bq')$. 
Integrating the both sides for $\bq$ with the 
weighting factor $\cos( q\wedge P)$, and 
regulating the resulting integral by the momentum 
cut off $\Lambda$, 
give an eigenvalue equation for 
the energy
\begin{eqnarray}
1=-{v_0\over 8\pi}
\left[ Z\left(-i\Lambda/\sqrt{\epsilon_B}\right)+ K_0 
\left(\theta P \sqrt{\epsilon_B} \right)\right]\,.
\label{eigenvalue}
\end{eqnarray} 
 This 
condition is precisely the one obtained in the 
perturbation theory and the  renormalization
may be achieved by the same way 
as in the perturbative analysis. 
The bound state 
energy is determined uniquely as a function of the renormalized 
coupling constant, the renormalization scale $\mu$ and the 
external momentum. 
For small $\theta P$, it can be explicitly 
solved  by
\begin{eqnarray}
E_r=-\left({2\mu\over \gamma\theta P}\right)^{1\over 2}\,\, 
e^{{8\pi\over v(\mu)}}\,.
\label{boundenergy}
\end{eqnarray} 
From (\ref{relative}), one may get the explicit form of the position 
space wavefunction,
\begin{eqnarray}
\phi =e^{i{\bP}\cdot {\bf R}}
\left[K_0\left(\sqrt{\epsilon_B}\, |\bx-{\!\!\!\phantom{1}_1\over 
\!\!\!\phantom{1}^2}\theta \tilde\bP|\right)+ 
K_0\left(\sqrt{\epsilon_B}\, 
|\bx+{\!\!\!\phantom{1}_1\over 
\!\!\!\phantom{1}^2}\theta \tilde\bP|\right)\right]\,,
\label{boundenergya}
\end{eqnarray} 
where $\tilde{P}^i=\epsilon^{ij}P_j$, ${\bf R}={\br_1\!+\!\br_2\over 2}$
and the relative position $\bx$ denotes $\br_1\!-\!\br_2$. 
The wavefunction has two 
distinguished  centers at $\pm {1\over 2}\theta \tilde\bP$
in the relative coordinates. 
The separation 
of these centers 
is transverse to the direction of the total 
momentum $P$. Furthermore, the separation grows linearly in the 
noncommutativity scale as well as  the total momentum.
This behavior is precisely the one expected in the open 
string theory in the constant NS-NS two form 
background\cite{susskind}, 
signalling the stringy nature of the field theories 
in noncommutative spacetime.

We now turn to the scattering problem. In the 
time independent momentum space equation in (\ref{twobodyb}), the 
scattering solution can be found with $E_r=k^2$ with an initial
relative momentum $\bk$. The exact scattering solution takes
a form
\begin{eqnarray}
\varphi(\bP,\bq)=(2\pi)^2\delta(\bq-\bk)
+C(\bP) {\cos (q\wedge P)\over q^2-k^2-i\epsilon} \,,
\label{scatteringsol}
\end{eqnarray}     
where $C(\bP)$ is as given above and $-i\epsilon$ is added
to select out the retarded Green function. Since $C(\bP)$ is 
dependent upon $\varphi$, it should be fixed self 
consistently. We again integrate the both sides of
 the above equation with respect to $\bq$ with weighting factor 
$\cos(q\wedge P)$ and obtain 
\begin{eqnarray}
C(\bP)=-{1\over 2}\,\,
{\cos(k\wedge P)  \over 
{1\over v_0}+{1\over 8\pi}
[ Z(\Lambda/k)+ 
K_0 (-i\theta k P)]} \,.
\label{cp}
\end{eqnarray}   
The exact scattering wavefunction is then
obtained from (\ref{scatteringsol}) as
\begin{eqnarray}
\varphi(\br)=e^{i\bk\cdot \br}+ {i\over 8}
\left[H^{(1)}_0\left(k |\br-{\!\!\!\phantom{1}_1\over 
\!\!\!\phantom{1}^2}\theta \tilde\bP|\right)+
H^{(1)}_0\left(k |\br+ 
{\!\!\!\phantom{1}_1\over 
\!\!\!\phantom{1}^2}\theta \tilde\bP|\right) \right]C(\bP)\,.
\label{scatteringsolb}
\end{eqnarray} 
where $H^{(1)}_\nu (x)$ is the Hankel function of the  first kind. 
The wave function 
takes a form in the asymptotic region as
\begin{eqnarray}
\varphi(\br)=e^{i\bk\cdot \br}+ {1\over \sqrt{r}}f(\alpha)
e^{i(k r+{\pi\over 4})}\,.
\label{scattering}
\end{eqnarray} 
with the scattering amplitude
\begin{eqnarray}
f(\alpha)= -{1\over 4\sqrt{\pi k}}\,\,{
\cos(k\wedge P) 
\cos({k'}\wedge P) 
\over {1\over v_0}+{1\over 8\pi}
[ Z(\Lambda/k)+ 
K_0 (-i\theta k P)]}\,,
\label{scatteringamp}
\end{eqnarray} 
where $k'_i = k \hat{r}_i$ 
and  $\alpha$ is the scattering angle 
between the initial momentum $\bk$ and the observation 
direction 
$\hat{r}$\footnote{Since we are dealing with identical 
particles, the wave function 
in (\ref{scatteringsolb}) should be 
symmetrized under $\br\!\rightarrow\! -\br$. This effect is added 
to (\ref{scatteringamp}) by the factor $\sqrt{2}$.}. 
The renormalization is again completed redefining the 
coupling constant as (\ref{renormalization}) while replacing
$\Lambda$ by $\mu$, and the scattering amplitude perfectly 
agrees with the field theoretic result. The wave function has again two 
centers, whose separation grows linearly with $\theta P$ to 
the transverse direction of the total momentum.
We note that the second cosine factor in
the scattering amplitude originates from the separation of the two 
centers and hence this peculiar behavior is even essential in obtaining
the correct scattering amplitude.

In this note, we have verified that the theory is 
renormalizable in the two particle sectors. In the 1PI 
$n\,(\geq\! 6 )$-point
Green function, the relevant loop integrals are UV 
finite. Thus the theory is expected to be 
renormalizable for any sectors. We leave the details 
for future works. Finally, we remark that our model may be 
relevant to studying the light-cone description of four 
dimensional relativistic field theory with $\theta^{\mu\nu}$
in the transverse directions only. 
This is because the model is reduced to a 
nonrelativistic theory on a noncommutative plane though
the interactions are typically more general than the one 
considered here.


There are many directions to go further. If one considers a fermionic 
nonrelativistic field on an ordinary plane, 
the contact interaction term vanishes identically 
due to the anticommuting 
nature of the fermionic field. On the noncommutative plane, 
the fermionic 
contact interaction becomes nontrivial and clarifications  are
required especially for the nature of the wavefunctions.
Rather trivial generalization of the system will include
$n$ component generalization or matrix Schr\"odinger field,
and the role of the contact interactions
will also be of interest. 
Another is to couple gauge fields to the noncommutative 
system. One of the simple option for this direction
is to couple the Chern-Simons gauge field, which describes
Aharonov-Bohm interaction between particles in case of the
 ordinary 
field theory\cite{bergman2}. The effect of the noncommutativity 
on the
Aharonov-Bohm interaction requires further studies.

\end{document}